\documentstyle[epsfig,rotate,preprint,floats,tighten,aps,amssymb]{revtex}

\begin{document}
\draft

\preprint{\vbox{\noindent\null \hfill ADP-99-52/T388\\ 
                         \null \hfill hep-lat/9912052 \\
}}

\title{\huge A Lattice QCD Analysis of the Strangeness \\ Magnetic Moment
of the Nucleon}

\author{Derek B. Leinweber\footnote{dleinweb@physics.adelaide.edu.au} and 
        Anthony W. Thomas\footnote{athomas@physics.adelaide.edu.au}}

\address{Department of Physics and Mathematical Physics\break
         and
         Special Research Centre for the Subatomic Structure of Matter,\break
         University of Adelaide, Australia 5005}
\maketitle

\begin{abstract}
The outcome of the SAMPLE Experiment suggests that the strange-quark
contribution to the nucleon magnetic moment, $G_M^s(0)$, may be
greater than zero.  This result is very difficult to reconcile with
expectations based on the successful baryon magnetic-moment
phenomenology of the constituent quark model.  We show that careful
consideration of chiral symmetry reveals some rather unexpected
properties of QCD.  In particular, it is found that the valence
$u$-quark contribution to the magnetic moment of the neutron can
differ by more than 50\% from its contribution to the $\Xi^0$ magnetic
moment.  This hitherto unforeseen result leads to the value $G_M^s(0)
= -0.16 \pm 0.18$ with a systematic error, arising from the relatively
large strange quark mass used in existing lattice calculations, that
would tend to shift $G_M^s(0)$ towards small positive values.
\end{abstract}

%\vspace{1.0cm}
%\indent PACS: 21.10.Ky, 12.39.Ba, 12.38.Gc, 11.30.Rd 

\newpage

\section{Introduction}

The SAMPLE collaboration recently reported\cite{SAMPLE} a new
experimental measurement of the strange-quark contribution to the
nucleon magnetic moment, $G_M^s(0)$:
\begin{equation}
G_M^s(0.1\ {\rm GeV}^2) = +0.61 \pm 0.17 \pm 0.21 \mu_N \, ,
\end{equation}
We note that the convention for $G_M^s$ is that the negative charge of
the strange-quark is not included in the definition. While the
uncertainties of the measurement and its interpretation (especially
the radiative corrections\cite{HOL}) are somewhat large and do not
totally exclude negative values, this result suggests that $G_M^s(0)$
is quite likely positive.  Prior to the appearance of the SAMPLE
result most theoretical analyses suggested that $G_M^s(0)$ should be
significantly below zero.

An analysis of the result of the SAMPLE experiment within the
framework of lattice QCD \cite{Leinweber:1996ie}, led to the
conclusion that a positive value for $G_M^s(0)$ is extremely difficult
to reconcile with our intuitive expectations based on the constituent
quark model.  In the light of the confirmation of the earlier SAMPLE
result, we have been led to reconsider this analysis taking into
account the recent progress in incorporating chiral corrections into
the extrapolation of quantities such as magnetic moments and masses
calculated on the lattice \cite{Leinweber:1998ej,Thomas:1999mv,%
Leinweber:1999ig,Leinweber:1999bv,Thomas:1999ae}.  We find that once
the correct chiral behavior of lattice quantities is taken into
account $G_M^s(0)$ tends to be small and may even be slightly
positive.  However, in contrast with naive expectations, the effective
magnetic moments of the quarks are no longer universal quantities but
show a significant dependence on the hadronic environment.

Our investigation begins with the derivation of magnetic moment sum
rules for the contributions to octet-baryon magnetic moments from
topologically distinct diagrams arising in lattice QCD.  These sum
rules are {\em exact} consequences of charge symmetry (i.e., when the
light $u$ and $d$ current quark masses are equal and electromagnetic
corrections negligible). As charge symmetry has been shown to be
extremely reliable for strongly interacting systems \cite{ChargeSymm},
we believe this is a very good approximation. Next we focus on chiral
symmetry and the consequent nonanalytic dependence of the baryon
magnetic moments on the mass of the light quarks.  In particular, we
show that the pion cloud contribution enhances the component of the
magnetic moment of the neutron associated with the $u$-quark by 50\%
compared with that associated with the $u$-quark in the $\Xi^0$; a
substantial departure from the common expectation of equal
contributions.  As we will see, this is precisely the condition
required to accommodate $G_M^s(0) > 0$.

The outline of this paper is as follows.  Section \ref{sec:qs} briefly
reviews the principles behind the derivation of the lattice QCD
magnetic moment sum rules\cite{Leinweber:1996ie} and explores the
numerical consequences, especially the parameter regions within which
$G_M^s$ can be positive. In section \ref{sec:chiral} we examine the
chiral behavior of the quantities appearing in the sum rules,
including their effects on the numerical results. In particular, we
show that the effect of the chiral corrections is to move the lattice
predictions in precisely the direction required to make $G_M^s(0)$
small in magnitude and possibly positive; our best estimate is
$G_M^s(0) = -0.16 \pm 0.18$. In the final section we discuss the
results and make some concluding remarks.

\section{QCD Constraints on Contributions to Baryon Magnetic Moments}
\label{sec:qs}

\subsection{Valence Versus Sea Quarks in Baryons}

\begin{figure}[b]
\vspace{-1.5cm}
\begin{center}
\setlength{\unitlength}{1.0cm}
\setlength{\fboxsep}{0cm}
\begin{picture}(10,5)
\put(0,0){\begin{picture}(5,5)\put(0,0){
\rotate{\epsfig{file=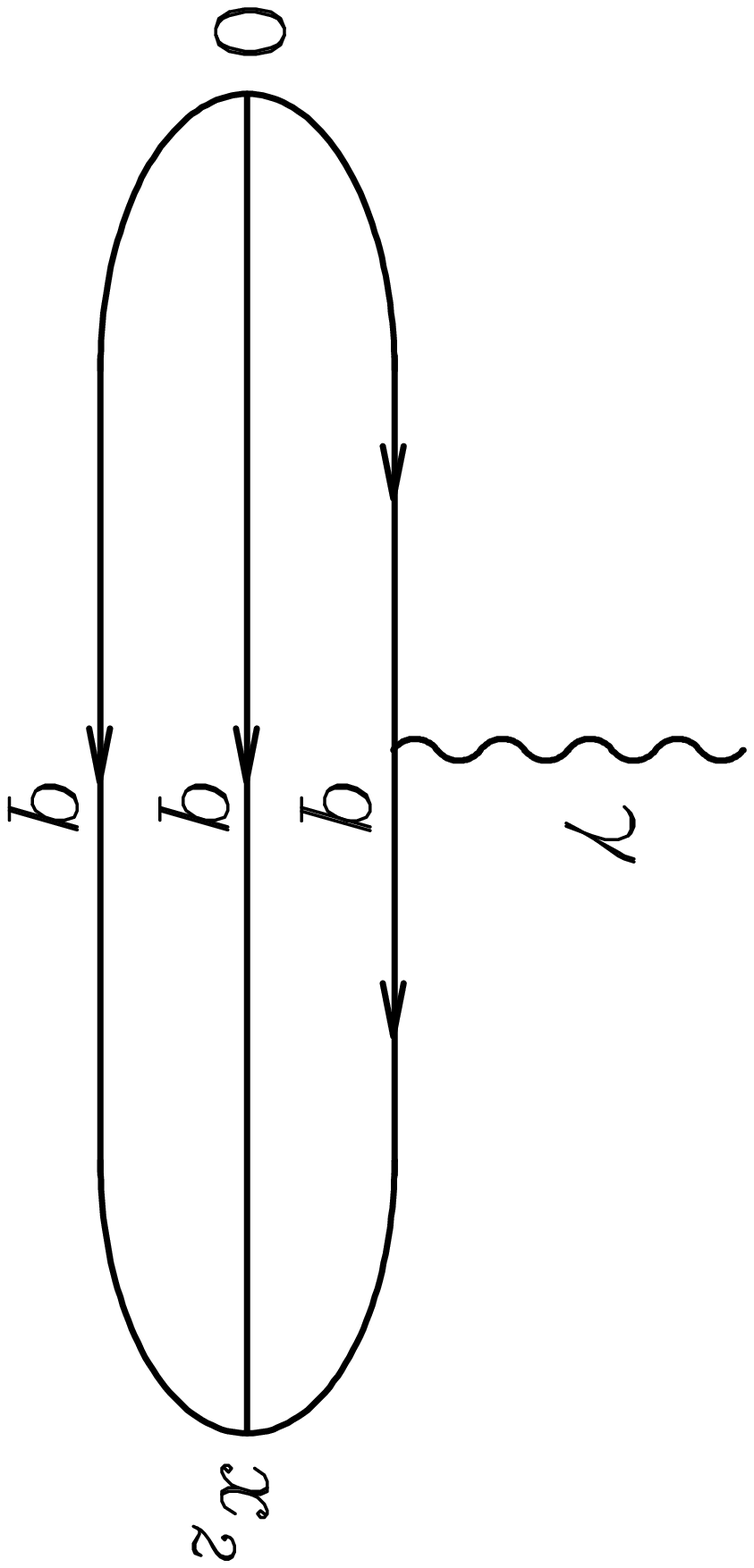,height=4cm}}}\end{picture}}
\put(5.5,0){\begin{picture}(5,5)\put(0,0){
\rotate{\epsfig{file=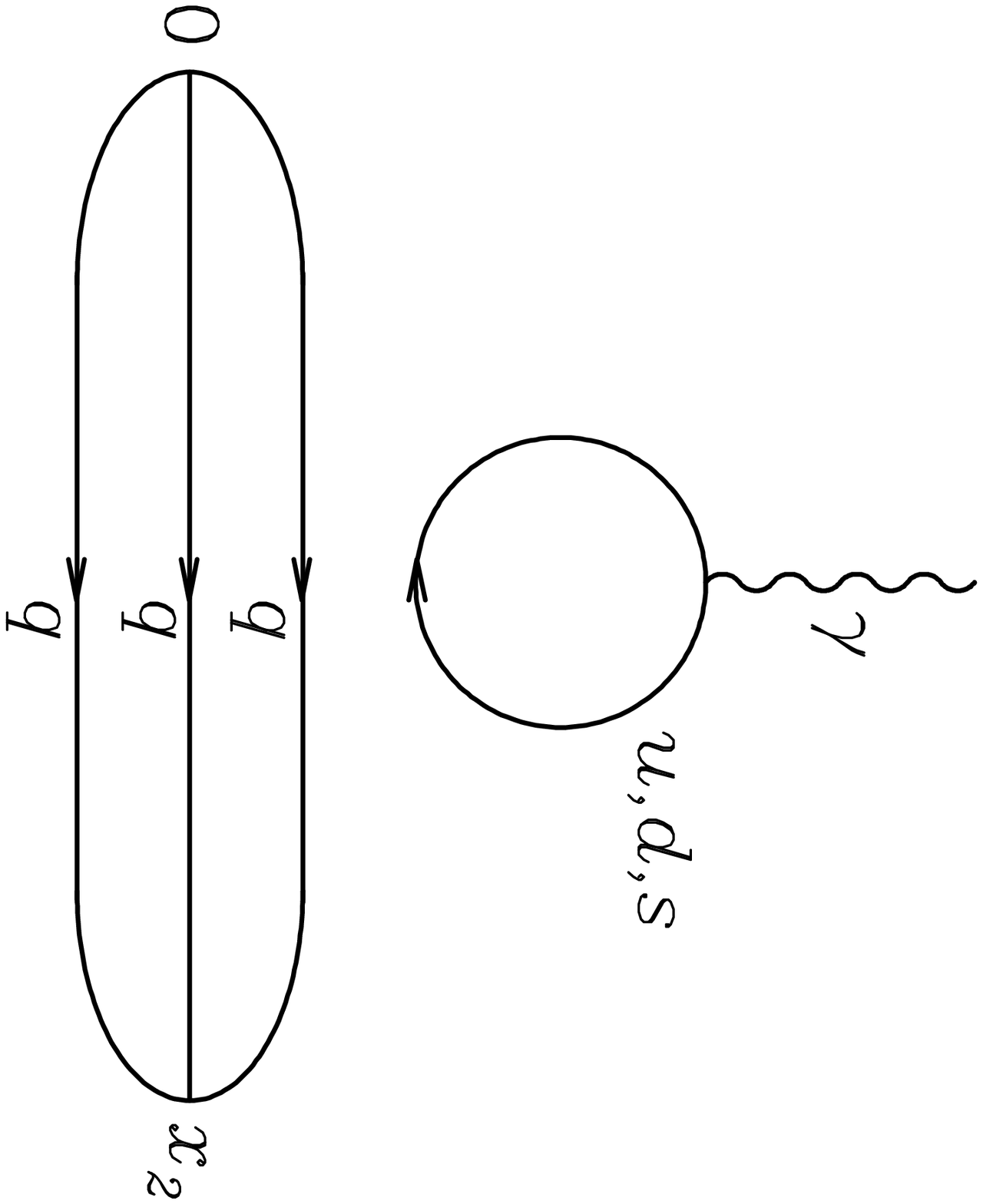,height=4cm}}}\end{picture}}
\end{picture}
\end{center}
\caption{Diagrams illustrating the two topologically different
insertions of the current within the framework of lattice QCD.  
These skeleton diagrams for the connected
(left) and disconnected (right) current insertions may be dressed by
an arbitrary number of gluons and quark loops.}
\label{topology}
\end{figure}

The Euclidean path integral formulation of quantum field theory is the
origin of fundamental approaches to the study of quantum
chromodynamics (QCD) in the nonperturbative regime.  An examination of
the symmetries manifest in the QCD path integral for current matrix
elements reveals various relationships among the quark sector
contributions \cite{Leinweber:1996ie}.  As we will see, these
relationships, together with the constraints of charge symmetry (which
should be very accurate) are sufficient to express the strange-quark
contribution to the nucleon magnetic moment, $G_M^s(0)$, in terms of
the experimentally measured baryon magnetic moments of $p$, $n$,
$\Sigma^+$, $\Sigma^-$, $\Xi^0$ and $\Xi^-$, and two ratios of quark
contributions to magnetic moments.

We begin by clearly defining the nature of the various quark sector
contributions referred to throughout this manuscript. We stress that
our theoretical analysis is made within the framework of full
(unquenched) QCD, so the line diagrams should be understood as
including any number of quark loops and gluons. For our purposes the
most important division to define is that between the ``sea''-quark
and ``valence''-quark contributions to the magnetic moment of a
baryon.

Current matrix elements of hadrons, such as the magnetic moment of the
proton, are extracted from the three-point function, a time-ordered
product of three operators\cite{Leinweber:1996ie}.  Generally, an
operator exciting the hadron of interest from the QCD vacuum is
followed by the current of interest, which in turn is followed by an
operator annihilating the hadron back to the QCD vacuum.  In
calculating the three point function, one encounters two topologically
different ways of performing the electromagnetic current insertion.
Figure \ref{topology} displays skeleton diagrams for these two
possible insertions (with Euclidean time increasing to the right).
(As already noted these diagrams may be dressed with an arbitrary
number of gluons and quark loops.)  The left-hand diagram illustrates
the connected insertion of the current to one of the ``valence''%
\footnote{Note that the term ``valence'' used here differs from that
commonly used in the framework of deep-inelastic scattering.
Here ``valence'' simply describes the quark whose quark
flow line runs continuously from $0 \to x_2$.  These lines can flow
backwards as well as forwards in time and therefore have a sea
contribution associated with them \protect\cite{dblPiCloud}.}
quarks of the baryon.  In the right-hand diagram the external field
produces a $q \, \overline q$ pair which in turn interacts with the
valence quarks of the baryon via gluons. It is important to realize
that within the lattice QCD calculation of the loop diagram on the
right in Fig. 1 there is no antisymmetrization (Pauli blocking) of the
quark in the loop with the valence quarks. For this reason, in general
{\em only the sum of the two processes in Fig. 1 is physical}.

The ratios of contributions to the magnetic moment of the nucleon
which are directly relevant to the determination of the strangeness
magnetic moment include the ratio of the loops (right-hand side of
Fig. 1) involving $s$ and $d$ quarks and one of two ratios of valence
quark contributions (left hand side of Fig. 1); either the ratio
$u_p/u_{\Sigma^+}$ (expressing the ratio of the valence $u$-quark
contribution in the proton to the analogous contribution when the
$u$-quark resides in the $\Sigma$ baryon), or the ratio
$u_n/u_{\Xi^0}$ (expressing the contribution of a valence $u$-quark to
the neutron relative to that when the $u$ quark resides in the
$\Xi^0$). Within simple constituent quark models, the $s/d$ quark loop
ratio is given by a ratio of constituent quark masses
\cite{Leinweber:1996ie} to be of order 0.65, while the valence-quark
ratios of interest here are equal to unity (i.e., $u_p/u_{\Sigma^+} =
u_n/u_{\Xi^0} = 1$).

\subsection{QCD Equalities}
\label{sec:eq}

Under the assumption of charge symmetry (namely equal mass light
quarks, $m_u=m_d$, and negligible electromagnetic corrections), the
three-point correlation functions for octet baryons leads to the
following equalities for electromagnetic current matrix elements
\cite{Leinweber:1996ie}:
\begin{eqnarray}
p =& e_u\, u_p + e_d\, d_p + O_N  \, , \qquad
n &= e_d\, u_p + e_u\, d_p + O_N  \, ,  \nonumber \\
\Sigma^+ =& e_u\, u_{\Sigma^+} + e_s\, s_\Sigma + O_\Sigma  \, , \qquad
\Sigma^- &= e_d\, u_{\Sigma^+} + e_s\, s_\Sigma + O_\Sigma  \, , \nonumber  \\
\Xi^0 =& e_s\, s_\Xi + e_u\, u_{\Xi^0} + O_\Xi  \, ,  \qquad
\Xi^- &= e_s\, s_\Xi + e_d\, u_{\Xi^0} + O_\Xi  \, .
\label{equalities}
\end{eqnarray}
Here, $O$ denotes the contributions from the quark-loop sector --
shown on the right-hand side of Fig. \ref{topology}.  The baryon label
represents the magnetic moment.  Subscripts allow for environment
sensitivity implicit in the three-point function
\cite{Leinweber:1996ie}.  For example, the three-point function for
$\Sigma^+$ is the same as for the proton, but with $d$ replaced by
$s$.  Hence, the $u$-quark propagators in the $\Sigma^+$ are
multiplied by an $s$-quark propagator, whereas in the proton the
$u$-quark propagators are multiplied by a $d$-quark propagator.  The
different mass of the neighboring quark gives rise to an environment
sensitivity in the $u$-quark contributions to observables, which means
that the naive expectations $u_p/u_{\Sigma^+} = u_n(\equiv
d_p)/u_{\Xi^0} = 1$ may not be satisfied
\cite{Leinweber:1996ie,dblOctet,dblMagMomSR,dblDecuplet,%
dblDiquarks,dblShedLight,dblEssential}.  This observation should be
contrasted with the common assumption that the quark magnetic moment
is an intrinsic-quark property which is independent of the quark's
environment.

Focusing now on the nucleon, we note that for magnetic properties,
$O_N$ contains sea-quark-loop contributions from primarily $u$, $d$,
and $s$ quarks.  In the SU(3)-flavor limit ($m_u = m_d = m_s$) the
charges add to zero and hence the sum vanishes.  However, the heavier
strange quark mass allows for a result which is non-zero.  By
definition
\begin{eqnarray}
O_N &=& {2 \over 3} \,{}^{\ell}G_M^u - {1 \over 3} \,{}^{\ell}G_M^d -
{1 \over 3} \,{}^{\ell}G_M^s \, , \\
&=& {\,{}^{\ell}G_M^s \over 3} \left ( {1 - \,{}^{\ell}R_d^s \over
\,{}^{\ell}R_d^s } \right ) \, , \quad \mbox{where} \quad
{}^{\ell}R_d^s \equiv {\,{}^{\ell}G_M^s \over \,{}^{\ell}G_M^d} \, ,
\label{OGMs}
\end{eqnarray}
and the leading superscript, $\ell$, reminds the reader that the
contributions are loop contributions.  Note that, in deriving
Eq.(\ref{OGMs}), we have set ${}^{\ell}G_M^u = {}^{\ell}G_M^d$,
corresponding to $m_u = m_d$ \cite{Leinweber:1996ie}.  In the
constituent quark model $\,{}^{\ell}R_d^s = m_d/m_s \simeq 0.65$.
However, we will consider $\,{}^{\ell}R_d^s$ in the range $-2$ to 2.

With no more than a little accounting, the quark-loop contributions to
the nucleon magnetic moment, $O_N$ may be isolated from
(\ref{equalities}) in the following two phenomenologically useful
forms,
\begin{eqnarray}
O_N &=& {1 \over 3} \left \{ 2\, p + n - {u_p \over u_{\Sigma^+}} \left (
\Sigma^+ - \Sigma^- \right ) \right \} \, , \label{disconnN1}  \\
O_N &=& {1 \over 3} \left \{ p + 2\, n - {u_n \over u_{\Xi^0}} \left (
\Xi^0 - \Xi^- \right ) \right \} \, . \label{disconnN2}  
\end{eqnarray}
As we explained above, under the assumption that quark magnetic
moments are not environment dependent, these ratios (i.e.
$\frac{u_p}{u_{\Sigma^+}}$ and $\frac{u_n}{u_{\Xi^0}}$) are taken to
be one in many quark models.  However, in order to explore the
validity of this assumption we will consider the range 0 to 2.
Equating (\ref{OGMs}) with (\ref{disconnN1}) or (\ref{disconnN2})
yields:
\begin{equation}
G_M^s = \left ( {\,{}^{\ell}R_d^s \over 1 - \,{}^{\ell}R_d^s } \right ) \left [
2 p + n - {u_p \over u_{\Sigma^+}} \left ( \Sigma^+ - \Sigma^- \right
) \right ] \, ,
\end{equation}
and
\begin{equation}
G_M^s = \left ( {\,{}^{\ell}R_d^s \over 1 - \,{}^{\ell}R_d^s } \right ) \left [
p + 2n - {u_n \over u_{\Xi^0}} \left ( \Xi^0 - \Xi^- \right ) 
 \right ] \, .
\end{equation}
Incorporating the experimentally measured baryon moments leads to:
\begin{equation}
G_M^s = \left ( {\,{}^{\ell}R_d^s \over 1 - \,{}^{\ell}R_d^s } \right ) \left [
3.673 - {u_p \over u_{\Sigma^+}} \left ( 3.618 \right ) \right ] \, , 
\label{ok}
\end{equation}
and
\begin{equation}
G_M^s = \left ( {\,{}^{\ell}R_d^s \over 1 - \,{}^{\ell}R_d^s } \right ) \left [
-1.033 - {u_n \over u_{\Xi^0}} \left ( -0.599 \right ) \right ] \, ,
\label{great}
\end{equation}
where all moments are expressed in nuclear magnetons $(\mu_N)$. (Note
that the measured magnetic moments are all known sufficiently
accurately\cite{PDG} that the experimental errors play no role in our
subsequent analysis.)  We stress that {\em these expressions for
$G_M^s$ are exact consequences of QCD, under the assumption of charge
symmetry}. Equation (\ref{great}) provides a particularly favorable
case for the determination of $G_M^s$ with minimal dependence on the
valence-quark ratio.
%
%\begin{figure}[tbp]
%\begin{center}
%\epsfig{file=GsMpx0848Pos.ps,width=12cm}
%\end{center}
%\caption{The surface for $G_M^s(0)\ge 0$ determined by equation
%(\protect\ref{great}).  Values for $G_M^s(0)$ outside the range $0
%\le G_M^s(0) \le 1$ are constrained to these limits for clarity.}
%\label{GsMnxPos}
%\end{figure}
%

\subsection{Necessary Conditions for $G_M^s(0)$ to be Positive}
\label{sec:bizarre}

Here we investigate the conditions that must hold if $G_M^s(0)$ is to
be greater than zero.  Since the $s$ quark is heavier than the $d$
quark, it would be very unusual to find an enhancement in the
$s$-quark-loop moment relative to the $d$-quark-loop moment.  Such an
occurrence would place the loop-ratio ${}^{\ell}R_d^s$ greater than
one.  Likewise, the mass effect is not expected to change the sign of
the $s$-quark-loop moment relative to the $d$-quark-loop moment, but
rather simply suppress the $s$-quark-loop moment relative to the
$d$-quark-loop moment.  Hence, the region of interest along the $s/d$
quark-loop ratio axis is the range (0,1) and it would be totally
unexpected if the ratio were found to lie outside this range.

Similarly, the success of constituent quark models in explaining the
octet-baryon magnetic moments suggests that the ratio of valence
$u$-quark moments in the neutron to $\Xi^0$ should be of order 1.  In
other words, although the $u$ quark is in an environment of two $d$
quarks in the neutron and two $s$ quarks in $\Xi^0$, $u_n/u_{\Xi^0}$
is expected to be the order of one.  It is important to recall that
the main difference between the $n$ and $\Xi^0$ magnetic moments is
the difference between the magnetic moment contributions of $s$ versus
$d$ quarks.  Traditionally, environment effects are regarded as a
secondary effect in the total baryon magnetic moment and often it is
completely disregarded. We shall show that, particularly for the
minority valence quarks (e.g, the $u$ in the neutron), this
expectation fails when the effects of chiral symmetry are taken into
account.

\subsubsection{Relation involving $u_n/u_{\Xi^0}$}

\begin{figure}[tbp]
\begin{center}
\epsfig{file=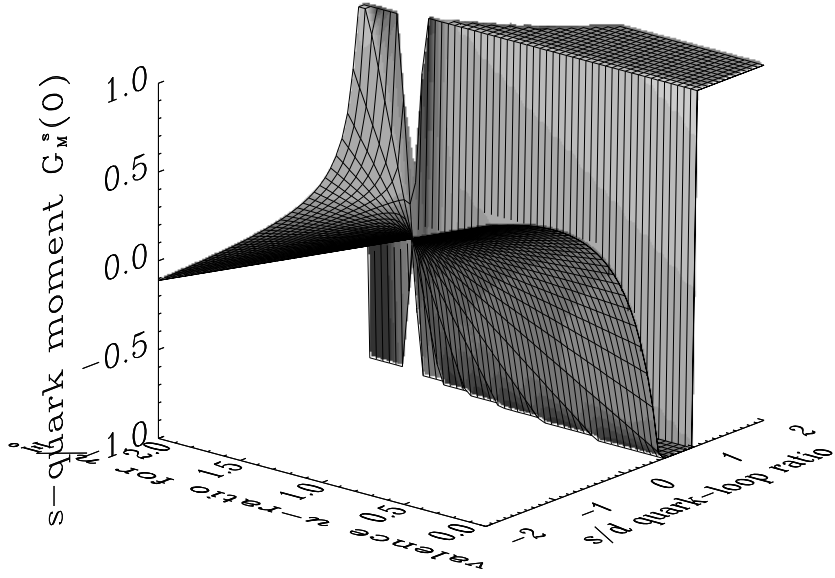,width=12cm}
\end{center}
\caption{The surface for $G_M^s(0)$ determined by equation
(\protect\ref{great}).  Values for $G_M^s(0)$ outside the range $-1
\le G_M^s(0) \le 1$ are constrained to these limits for clarity.}
\label{GsMnx}
\end{figure}

We now investigate the range of values of $u_n/u_{\Xi^0}$ and
${}^{\ell}R_d^s$ over which Eq.(\ref{great}) allows positive values of
$G_M^s(0)$. Figure \ref{GsMnx} illustrates the dependence of
$G_M^s(0)$ on these two ratios, based on Eq.(\ref{great}).

%\begin{figure}[tbp]
%\begin{center}
%\epsfig{file=GsMps1070Pos.ps,width=12cm}
%\end{center}
%\caption{The surface for $G_M^s(0)\ge 0$ determined by equation
%(\protect\ref{ok}).  Values for $G_M^s(0)$ outside the range $0
%\le G_M^s(0) \le 1$ are constrained to these limits for clarity.}
%\label{GsMpsPos}
%\end{figure}
%
%Figure \ref{GsMnxPos} illustrates the surface of (\ref{great}) where
%$G_M^s(0) \ge 0$.  We see that the solution of (\ref{great}) for the

Clearly, for the sea-quark-loop ratio, ${}^{\ell}R_d^s$, in the
expected range (0,1), the valence-quark moment ratio $u_n/u_{\Xi^0}$
must be very large, greater than 1.725. Linear extrapolations of
lattice QCD calculations (as a function of quark mass) for this ratio,
examining environment sensitivity, lead to a
ratio\cite{dblOctet,dblDecuplet} $u_n/u_{\Xi^0} = 0.72 \pm 0.46$.
This makes the large values required to stay in the region of interest
for the sea-quark-loop ratio look somewhat unusual.  Because of the
nature of the ratios involved, the systematic uncertainties in the
actual lattice QCD calculations are expected to be small relative to
the statistical uncertainties.  Hence the main source of uncertainty
is in the extrapolation of the lattice results to the physical world.

\subsubsection{Relation involving $u_p/u_{\Sigma^+}$}

\begin{figure}[tbp]
\begin{center}
\epsfig{file=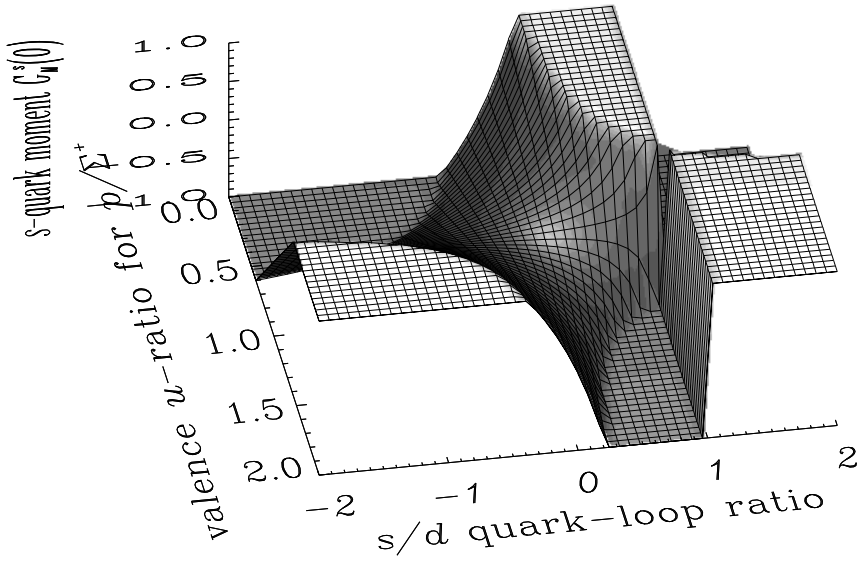,width=12cm}
\end{center}
\caption{The surface for $G_M^s(0)$ determined by equation
(\protect\ref{ok}).  Values for $G_M^s(0)$ outside the range $-1
\le G_M^s(0) \le 1$ are constrained to these limits for clarity.}
\label{GsMps}
\end{figure}

Here we study the dependence of $G_M^s(0)$ obtained from Eq.(\ref{ok})
as a function of the quark-sector ratios $u_p/u_{\Sigma^+}$ and
${}^{\ell}R_d^s$.  Figure \ref{GsMps} illustrates the value of
$G_M^s(0)$, based on (\ref{ok}), as a function of these two ratios.
%surface of (\ref{ok}) where $G_M^s(0) \ge 0$.  
We see that the solution of (\ref{ok}) for the sea-quark-loop ratio in
the region (0,1) requires the valence-quark moment ratio
$u_p/u_{\Sigma^+} < 1.015$.

\subsubsection{Consistency Relation}

\begin{figure}[tbp]
\centering{\
\rotate{\epsfig{file=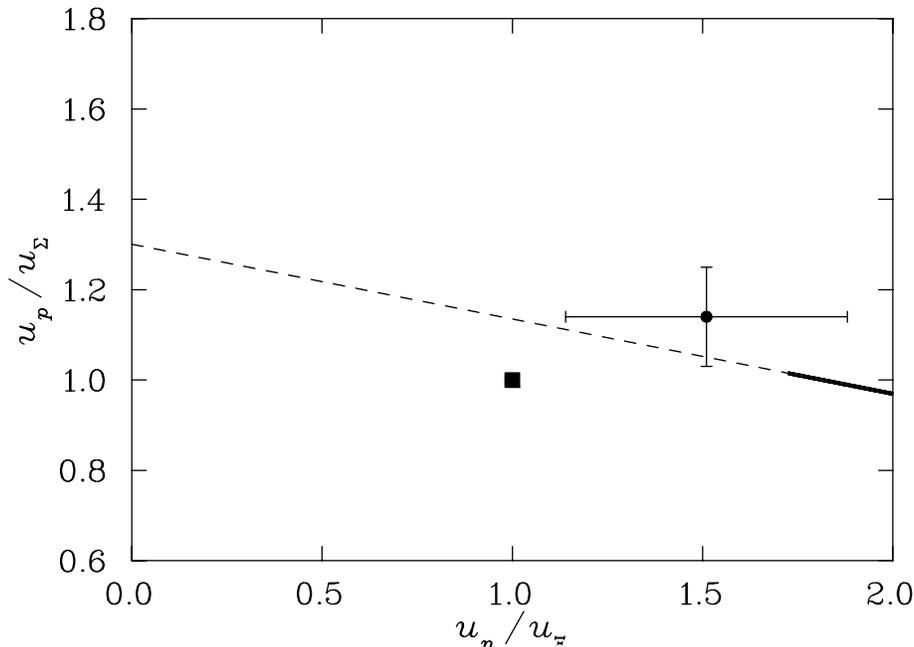,height=12cm}} }
\caption{The consistency relation between $u_p/u_{\Sigma^+}$ and
$u_n/u_{\Xi^0}$ which must be satisfied within QCD.  The part of the
straight line which is dashed corresponds to $G_M^s(0) < 0$, while the
solid part of the line has $G_M^s(0) > 0$.  The standard quark model
assumption of intrinsic quark moments independent of their environment
is indicated by the filled square.  The lattice QCD prediction (after
an appropriate chiral extrapolation, discussed in following sections)
is illustrated by the filled circle.  }
\label{SelfCons}
\end{figure}

Equating (\ref{ok}) and (\ref{great}) provides a linear relationship
between $u_p/u_{\Sigma^+}$ and $u_n/u_{\Xi^0}$ which must be satisfied
within QCD. Figure \ref{SelfCons} displays this relationship by the
dashed and solid line, the latter corresponding to values for which
$G_M^s(0) > 0$ when ${}^{\ell}R_d^s$ is in the anticipated range $0 <
{}^{\ell}R_d^s < 1$.  Since the line does not pass through the point
$(1.0, 1.0)$ corresponding to the simple quark model assumption, the
experimentally measured moments are signaling that there must be an
environment effect exceeding 12\% in both ratios or approaching 20\%
or more in at least one of the ratios.  Moreover, a positive value for
$G_M^s(0)$ requires an environment sensitivity exceeding 70\% in the
$u_n/u_{\Xi^0}$ ratio. 

\section{Chiral Corrections}
\label{sec:chiral}
 
One of the major challenges at present in connecting lattice
calculations of hadronic properties with the physical world is that
computational limitations restrict the accessible quark masses to
values an order of magnitude larger than the physical values. At such
large masses one is far from the region where chiral perturbation
theory is applicable.  Yet one knows that for current quark masses
near zero there is important non-analytic structure (as a function of
the quark mass) which must be treated correctly if we are to compare
with physical hadron properties. Our present analysis of the
strangeness magnetic form factor has been made possible by a recent
breakthrough in the treatment of these chiral corrections for the
nucleon magnetic moments \cite{Leinweber:1998ej,Thomas:1999mv}. In
particular, a study of the dependence of the nucleon magnetic moments
on the input current quark mass, within a chiral quark model which was
fitted to existing lattice data, suggested a model independent method
for extrapolating baryon magnetic moments which satisfied the chiral
constraints imposed by QCD.  We briefly summarize the main results of
that analysis:
\begin{itemize}
\item a series expansion of $\mu _{p(n)}$ in powers of $m_{\pi }$ is
not a valid approximation for $m_{\pi }$ larger than the physical
mass,
\item on the other hand, the behavior of the model, after adjustments
to fit the lattice data at large $m_{\pi }$ is well determined by
the simple Pad\'e approximant:
\begin{equation}\label{mag-mom}
\mu _{p(n)}=\frac{\mu _{0}}{1-\frac{\chi }{\mu _{0}}m_{\pi }+ c\,
m^{2}_{\pi }} .
\end{equation}

\item Eq.(\ref{mag-mom}) not only builds in the usual magnetic moment 
of a Dirac particle at
moderately large $m^{2}_{\pi }$ but has the correct leading non-analytic
(LNA) behavior of chiral perturbation theory
\[
\mu =\mu _{0} + \chi\, m_{\pi} ,
\]
with $\chi$ a model independent constant.
\item fixing $\chi$ at the value given by chiral perturbation theory
and adjusting $\mu _{0}$ and $c$ to fit the lattice data yielded
values of $\mu _{p}$ and $\mu _{n}$ of $2.85\pm 0.22\, \mu _{N}$ and
$-1.96\pm 0.16\, \mu _{N}$, respectively, at the physical pion
mass. These are in remarkably good agreement with the experimental
values -- certainly much closer than the usual linear extrapolations
in $m_{q}$.
\end{itemize}

Clearly it is vital to extend the lattice calculations of the baryon
magnetic moments to lower values of $m_{\pi }$ than the 600 MeV used
in the study just outlined.  It is also vital to include dynamical
quarks.  Nevertheless, the apparent success of the extrapolation
procedure gives us strong encouragement to investigate the same
approach for resolving the strange quark contribution to the proton
magnetic moment.

\subsection{Chiral Behavior of the Valence and Sea Contributions}

\begin{figure}[t]
\begin{center}
\epsfig{file=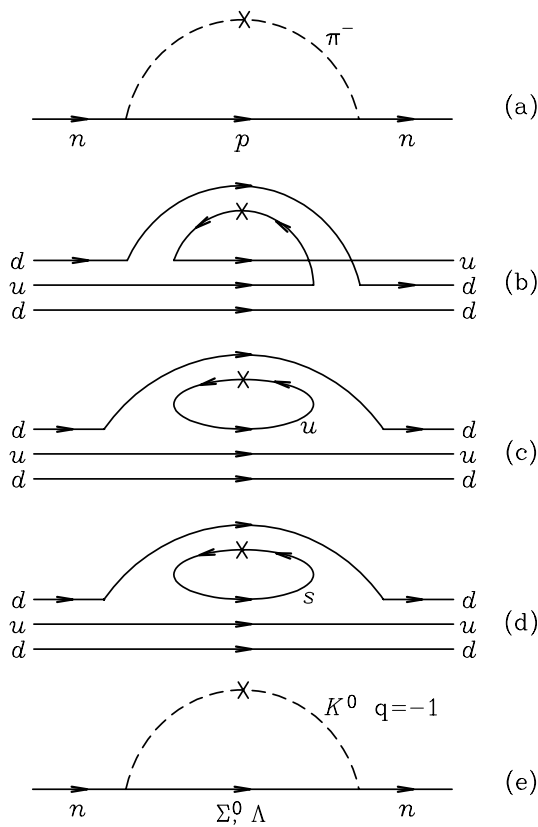,width=7cm}
\end{center}
\caption{The components of the leading non-analytic contribution to the
magnetic moment of the neutron -- (a) through (c), as well as related
processes, (d) and (e), with the same chiral behavior as (c).}
\label{UinNpim}
\end{figure}

The study of the chiral behavior of hadronic properties calculated on
the lattice is still at a relatively early stage of development. For
quenched QCD there has been a very thorough investigation of the
corrections to hadron masses \cite{SHARPE,Golterman}.  However, for
magnetic moments little is known apart from the constraints of chiral
perturbation theory for full QCD.  Fig.\ \ref{UinNpim}(a) illustrates
the process which gives rise to the leading non-analytic behavior of
the neutron magnetic moment.  (The cross represents the
electromagnetic current operator.)  The breakdown of this process into
lattice valence and sea contributions is shown in
Figs.\ref{UinNpim}(b) and (c), respectively.  Note that although
Fig.\ref{UinNpim}(b) involves propagation of one of the valence quarks
backward in time, its topology is equivalent to that of Fig. 1(a) and
it is included in it.

It is important to realize that the separation of Fig.\
\ref{UinNpim}(a) into the components (b) and (c), while perfectly
reasonable within a lattice formulation of the problem, is not
physical.  That is, because there is no antisymmetrization exchange
between the quark in the loop and the valence quarks in Fig.\
\ref{UinNpim}(c), the quantum numbers of the baryons appearing in the
intermediate states of Figs.\ref{UinNpim}(b) and (c) are not
necessarily allowed.  Only the sum of the two diagrams, in which the
unphysical contributions cancel, has a physical meaning.  On the other
hand, if one is to analyze the baryon magnetic moments within the
framework of lattice QCD, as outlined in Sect.2, one must understand
the chiral behavior of Figs.\ \ref{UinNpim}(b) and (c) separately.

Fortunately this is not so difficult. We know the leading non-analytic
behavior of the full diagram, Fig.\ \ref{UinNpim}(a), so that, if we
can calculate the corresponding behavior of either (b) or (c) we know
the other.  Fig.\ \ref{UinNpim}(c) involves a $u$-quark loop where
(within the calculational rules for lattice QCD) no exchange term is
possible.  Thus the $u$-quark in the loop is distinguishable from all
the other quarks in the diagram.  The chiral structure of this diagram
is therefore identical to that for a ``strange'' quark loop, as
illustrated in Fig.\ \ref{UinNpim}(d), provided the ``strange'' quark
appearing here is understood to have the same mass as the $u$-quark.

The corresponding hadron diagrams which give rise to the leading
non-analytic structure of Fig.\ \ref{UinNpim}(c) are therefore those
shown in Fig.\ref{UinNpim}(e), with the distinguishable (``strange'')
quark mass set equal to the mass of the $u$-quark.  That is, the
intermediate, spin-$\frac{1}{2}$ baryons appearing in
Fig.\ref{UinNpim}(e) should be degenerate with the nucleon.
Similarly, the ``kaon'' mass is degenerate with the pion.  In
addition, it should be reasonable to estimate the coupling constants
using SU(6) symmetry -- note that in this case all the quarks {\bf do}
have the same mass.

As a check of this technique for extracting the LNA behavior, we
confirmed that our results for the chiral contribution of the valence
term to the mass of the nucleon obtained with our technique agrees
with that obtained for quenched QCD, in the more formal approach of
Labrenz and Sharpe \cite{SHARPE} -- after allowing for the $\eta$ and
$\eta '$ contributions which are unique to the quenched case.  (Note
that these authors also use SU(6) coupling constants.)

The only residual technique needed is that, in order to pick out the
$u$-quark contribution to these various diagrams -- e.g., the quantity
${}^{\ell}G_M^u$ appearing in Eq. (\ref{OGMs}) or $u_n$ in
Eq. (\ref{great}) -- we set the electromagnetic charge of the
$u$-quark to be +1 in Fig.\ \ref{UinNpim}(b) and Fig.\
\ref{UinNpim}(c) and the $d$-quark charge to zero.  As the ``strange''
quark of Fig.\ \ref{UinNpim}(d) is accounting for the $u$-quark
contribution in Fig.\ \ref{UinNpim}(c), this ``strange'' quark also
takes charge $+1$.  This is the reason for the label $q=-1$ on the
``$K^0$'' in Fig.\ \ref{UinNpim}(e), which ensures that the
anti-$u$-quark in Fig.\ \ref{UinNpim}(c) is counted correctly.

\begin{figure}[t]
\begin{center}
\epsfig{file=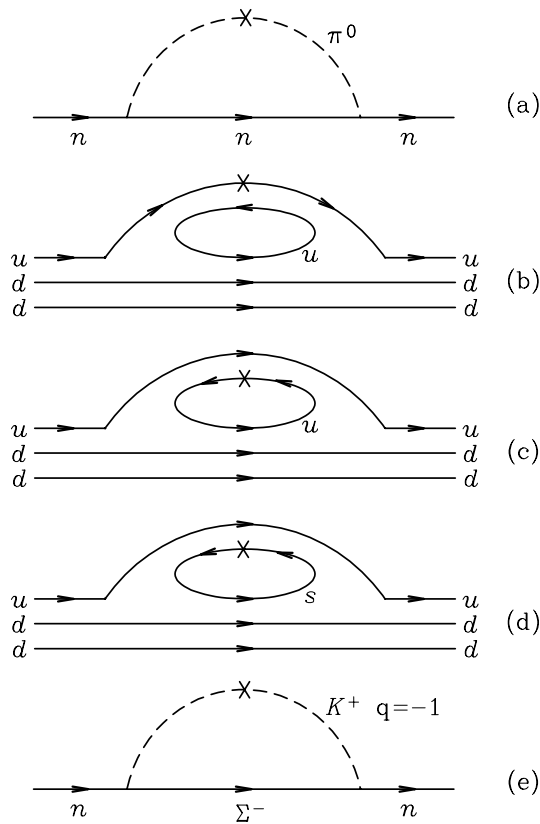,width=7cm}
\end{center}
\caption{Separation of valence and sea contributions to the $\pi^0$
loop contribution to the magnetic moment of the neutron -- (a) through
(c).  While the net effect is vanishing, valence and sea contributions
are equal and opposite in sign.  The related processes, (d) and (e),
with the same chiral behavior as (c) are discussed in the text.}
\label{UinNpi0}
\end{figure}

The separation of valence and sea $u$-quark contributions must also be
carried out for the process illustrated in Fig.\ \ref{UinNpi0}.  While
the net effect of the $\pi^0$ cloud is vanishing, valence and loop
contributions are equal and opposite in sign.  As in Fig.\
\ref{UinNpim}, Fig.\ \ref{UinNpi0}(c) involves a $u$-quark loop where
no exchange term is possible.  Thus the $u$-quark in the loop is
essentially distinguishable from all the other quarks in the diagram.
The chiral structure of this diagram is identical to that for a
strange quark loop, as illustrated in Fig.\ \ref{UinNpi0}(d), provided
the ``strange'' quark appearing here is understood to have the same
mass as the $u$-quark.  The corresponding hadron diagrams which give
rise to the leading non-analytic structure of Fig.\ \ref{UinNpi0}(c)
are therefore those shown in Fig.\ref{UinNpi0}(e), with the
distinguishable (``strange'') quark mass set equal to the mass of the
$u$-quark.  Again, we set the electromagnetic charge of the $u$-quark
to be +1 and the $d$-quark charge to zero in Fig.\ \ref{UinNpi0}(b)
and (c).  As the ``strange'' quark of Fig.\ \ref{UinNpi0}(d) is
accounting for the $u$-quark contribution in Fig.\ \ref{UinNpi0}(c),
this ``strange'' quark also takes charge $+1$ and the $u$ and $d$
charges are set to zero.  In this way the ``$K^+$'' has charge $q=-1$
in Fig.\ \ref{UinNpi0}(e), which ensures that the anti-$u$-quark in
Fig.\ \ref{UinNpi0}(c) is counted correctly.

A summary of the isospin and charge factors associated with the
leading non-analytic (LNA) chiral coefficients is given in Table
\ref{table:isocharge}.  The LNA pieces of these various contributions
to the baryon magnetic moments are proportional to
\begin{equation}
\beta \, {m_N \over 8 \, \pi \, f_\pi^2} \, m_\pi \equiv \chi \, m_\pi \, ,
\label{LNAdef}
\end{equation}
where the pion decay constant $f_\pi = 93$ MeV.  The coefficients
$\beta$ and $\chi$ for various quark sectors and baryons are
summarized in Table \ref{table:chiralCoeff}. (Note that $\beta$ and
$\chi$ are, of course, equivalent but for completeness we give
numerical values for both, as well as the analytic expression for
$\beta$ in terms of the SU(6) constants $F$ and $D$.) The heading
``$u$-flavor-sector'' is the $u$-quark contribution to the LNA
behavior of the baryon magnetic moments in full QCD; ``$u$-sea-quark
loop'' gives the terms we just discussed (such as Fig.\
\ref{UinNpim}(c) and Fig.\ \ref{UinNpi0}(c)) and ``$u$-valence
sector'' is the difference (corresponding, for example, to Fig.\
\ref{UinNpim}(b) plus Fig.\ \ref{UinNpi0}(b)).
\begin{table}[tb]
{\centering 
\begin{tabular}{lcc}
Baryon &$u$-flavor sector &$u$-sea-quark loop \\
\hline 
$n$        &$-2\, f_{\pi NN}^2$ 
           &$-(3\, f_{KN\Sigma}^2 + f_{KN\Lambda}^2)$\\
$\Xi^0$    &$+2\, f_{\pi \Xi \Xi}^2$ 
           &$-2\, f_{\pi \Xi \Xi}^2$\\
$p$        &$+2\, f_{\pi NN}^2$ 
           &$-(3\, f_{KN\Sigma}^2 + f_{KN\Lambda}^2)$\\
$\Sigma^+$ &$f_{\pi \Sigma \Sigma}^2 + f_{\pi \Sigma \Lambda}^2$
           &$-(f_{\pi \Sigma \Sigma}^2 + f_{\pi \Sigma \Lambda}^2)$ \\
\end{tabular}\par}
\caption{Isospin and charge factors for the leading nonanalytic
contribution of the $u$-quark(s) to the magnetic moment of various
baryons.  The $u$-quark charge has been normalized to 1, as explained in
the text.  The
contribution of the valence quark sector alone is obtained by
subtracting the sea-quark loop contribution 
(``$u$-sea-quark loop'') from the total $u$-flavor
sector (``$u$-flavor-sector'').
\label{table:isocharge}}
\end{table}
\begin{table}[tbp]
{\centering 
\begin{tabular}{lcccc}
Baryon &Coefficient &$u$-flavor sector &$u$-sea-quark loop
&$u$-valence sector \\ 
\hline 
$n$        &$\beta$     
           &$(F+D)^2$ 
           &$(9\, F^2 - 6\, FD + 5\, D^2)/3$
           &$2 (-3\, F^2 + 6\, FD - D^2)/3$ \\
           &$\beta$  &$1.020$ &$0.612$ &$0.408$ \\ 
           &$\chi$   &$4.41$  &$2.65$  &$1.76$  \\
$\Xi^0$    &$\beta$     
           &$-(F-D)^2$ 
           &$+(F-D)^2$ 
           &$-2(F-D)^2$ \\
           &$\beta$  &$-0.0441$ &$0.0441$ &$-0.0882$ \\ 
           &$\chi$   &$-0.191$  &$0.191$  &$-0.381$  \\
$p$        &$\beta$     
           &$-(F+D)^2$ 
           &$(9\, F^2 - 6\, FD + 5\, D^2)/3$
           &$-4 (3\, F^2 + 2\, D^2)/3$ \\
           &$\beta$  &$-1.020$ &$0.612$ &$-1.632$ \\ 
           &$\chi$   &$-4.41$  &$2.65$  &$-7.06$  \\
$\Sigma^+$ &$\beta$     
           &$-2 (3\, F^2 + D^2)/3$
           &$2 (3\, F^2 + D^2)/3$
           &$-4 (3\, F^2 + D^2)/3$ \\
           &$\beta$  &$-0.568$ &$0.568$ &$-1.136$ \\ 
           &$\chi$   &$-2.46$  &$2.46$  &$-4.91$  
\end{tabular}\par}
\caption{Coefficients $\beta$ and $\chi$ of (\protect\ref{LNAdef}) for
the leading nonanalytic (LNA) contribution of the $u$-quark(s) to the
magnetic moment of various baryons. We express the LNA coefficients
$\chi$ in terms of the usual SU(6) constants $F$ and $D$, as well as
giving numerical values for $\chi$ and $\beta$ (which are equivalent
through Eq.(\protect\ref{LNAdef})). The $u$-quark charge has been
normalized to 1, for reasons explained in the text.
\label{table:chiralCoeff}}
\end{table}

\begin{figure}[tbp]
\begin{center}
\epsfig{file=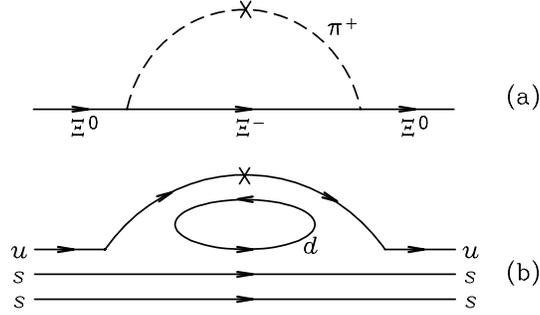,width=7cm}
\end{center}
\caption{The dominant leading non-analytic contribution to the
magnetic moment of the $\Xi^0$ -- see the text for the explanation of
process (b).}
\label{UinXi0pip}
\end{figure}

\begin{figure}[tbp]
\begin{center}
\epsfig{file=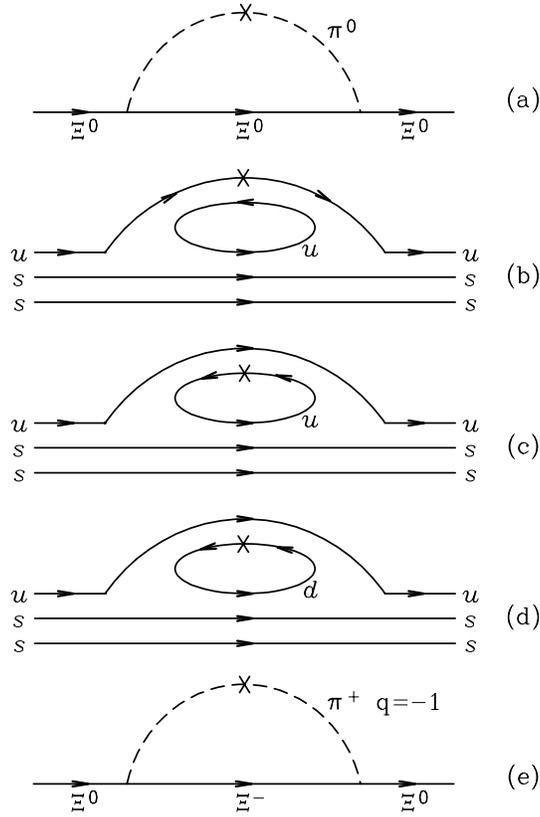,width=7cm}
\end{center}
\caption{Separation of valence and sea contributions to the $\pi^0$
loop contribution to the magnetic moment of $\Xi^0$ -- (a) through
(c).  The related processes, (d) and (e), with the same chiral
behavior as (c) are discussed in the text.}
\label{UinXi0pi0}
\end{figure}

For the calculation of $G_M^s(0)$, the LNA chiral behavior of $u_n$
and $u_{\Xi^0}$ is crucial, as the corresponding coefficients are of
opposite sign.  As a result, the ratio of these $u$-quark
contributions, extrapolated with the correct chiral behavior, will be
completely different from the linearly extrapolated ratio.  That the
signs of these chiral coefficients are opposite may be traced back to
the charge of the predominant pion cloud associated with these
baryons.  The chiral behavior in the neutron case is dominated by the
transition $n \to p\, \pi^- \to n$, where there is a contribution from
a $\bar{u}$ loop.  On the other hand, for the $\Xi^0$ the LNA
$u$-quark contribution is dominated by the process $\Xi^0 \to \Xi^-
\pi^+ \to \Xi^0$, shown in Fig.\ \ref{UinXi0pip}.  In this case there
is no contribution from a $u$-sea-quark loop, rather the virtual
$\pi^+$ involves a valence $u$-quark -- and hence the sign change.  Of
course, one must account for a $u$-quark loop contribution in the
process of Fig.\ \ref{UinXi0pi0}, which is completely analogous to
Fig.\ \ref{UinNpi0}.

In Fig.\ \ref{UinXi0pi0}(c) we have a $u$-quark loop where, according
to the rules of lattice field theory, no exchange term is possible.
Thus the $u$-quark in the loop is essentially distinguishable from all
the other quarks in the diagram.  The chiral structure of this diagram
is therefore identical to that for a $d$-quark loop, as illustrated in
Fig.\ \ref{UinXi0pi0}(d) provided the ``$d$-quark'' appearing here has
the same mass as the $u$-quark. It is convenient to use the $d$-quark
in this case as it does not appear as one of the valence quark
flavors, just as the strange quark could be used earlier because it is
not a valence flavor of the neutron.

As a result of these arguments, the hadronic diagrams which give rise
to the leading non-analytic structure of Fig.\ \ref{UinXi0pi0}(c) are
therefore those shown in Fig.\ \ref{UinXi0pi0}(e).  The
electromagnetic charge of the $d$-quark in the loop of Fig.\
\ref{UinXi0pi0}(d), accounting for the $u$-quark contribution of Fig.\
\ref{UinXi0pi0}(c), is set to +1 and all the other quark charges are
zero.  In this way the ``$\pi^+$'' has charge $q=-1$ in Fig.\
\ref{UinXi0pi0}(e), which ensures that the anti-$u$-quark in Fig.\
\ref{UinXi0pi0}(c) is counted correctly.

\subsection{Numerical Results including Chiral Behavior}

The original lattice QCD studies of the magnetic moments of the octet
baryons \cite{dblOctet,dblMagMomSR,dblDecuplet} contain enough
information that one can readily extract the contributions associated
with each flavor of valence quark. We stress that our theoretical
analysis has been based on full QCD, whereas the existing lattice
calculations have all been made within quenched QCD. On the other
hand, the range of quark masses over which these calculations have
been carried out corresponds to pion masses greater than 600 MeV.  The
study of Refs.\ \cite{Leinweber:1998ej,Thomas:1999mv} suggests that at
such large masses the errors arising from quenching should be less
than 10\%, since the {\em total} pion contribution is of this order.
Thus, while it would be highly desirable to have full QCD data, we do
not expect the errors arising from using the existing quenched data to
be unreasonable.

In order to extrapolate the lattice data for valence quarks to the
physical pion mass we use the same approach which worked so well for
the neutron and proton moments in Ref.\
\cite{Leinweber:1998ej,Thomas:1999mv}.  That is, we fit the lattice
data for the valence quark of flavor $q$ in baryon $B$ with the form:
\begin{equation}
\mu(m_\pi) = \frac{\mu^{0}}
{1 - \frac{\chi}{\mu^0} \, m_\pi + c \, m_\pi^2}.
\label{eq:chiral}
\end{equation}
Here $\chi$ are the modified LNA chiral coefficients for the valence
quarks given in Table II (last column) and $\mu^0$ and $c$ are fitting
parameters.  In the case of the $u$-quark in $\Xi^0$, the sign of
$\chi$ is such that a singularity is introduced in (\ref{eq:chiral});
a feature not present in the standard chiral expansion
\begin{equation}
\mu(m_\pi) \simeq \mu^{0} + \chi \, m_\pi + c \, m_\pi^2 \, .
\label{eq:stdchiral}
\end{equation}
Fortunately the coefficient $\chi$ is small and (\ref{eq:stdchiral})
should be adequate.  We have tested the sensitivity of the truncation
of the chiral expansion of (\ref{eq:stdchiral}) by also considering
the form
\begin{equation}
\mu(m_\pi) = \frac{\mu^{0} + \chi \, m_\pi}{1 + c \, m_\pi^3}.
\label{eq:altchiral}
\end{equation}
which has the correct chiral and heavy quark limits.  Extrapolations
based on this form agree with (\ref{eq:stdchiral}) at the 7 \%
level.  We present extrapolations based on (\ref{eq:stdchiral}) in the
following. 

\begin{figure}[tbp]
\centering{\
\rotate{\epsfig{file=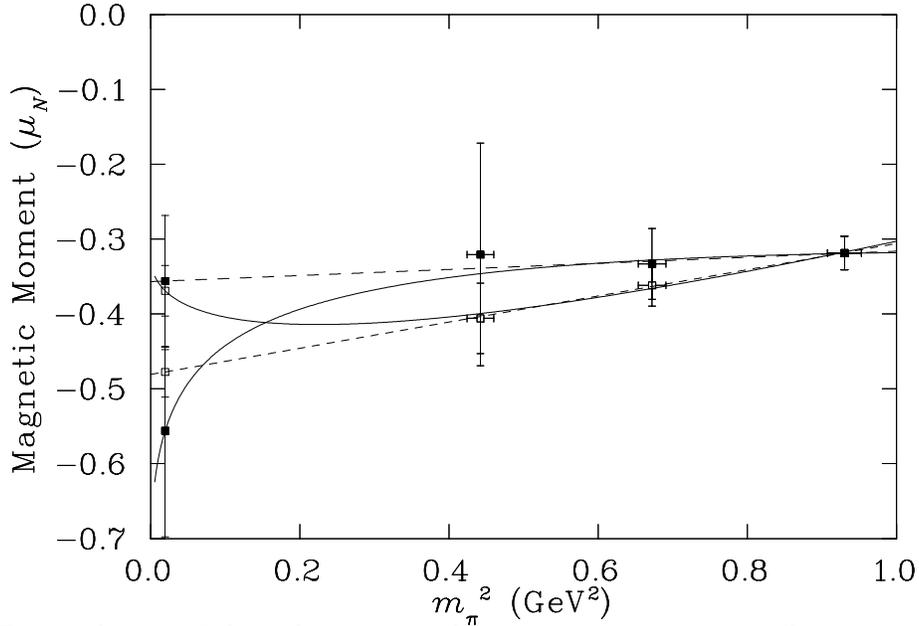,height=12cm}} }
\caption{Extrapolation of the valence $u$-quark magnetic moment
contributions to the neutron (solid symbols) and $\Xi^0$ (open
symbols).  Naive linear extrapolations (dashed lines) are contrasted
with the chiral extrapolations based on (\protect\ref{eq:chiral}) for
the neutron and (\protect\ref{eq:stdchiral}) for $\Xi^0$.
The lattice data is taken from 
Ref.\protect\cite{dblOctet}. 
}
\label{NXiExtrap}
\end{figure}

\begin{figure}[tbp]
\centering{\
\rotate{\epsfig{file=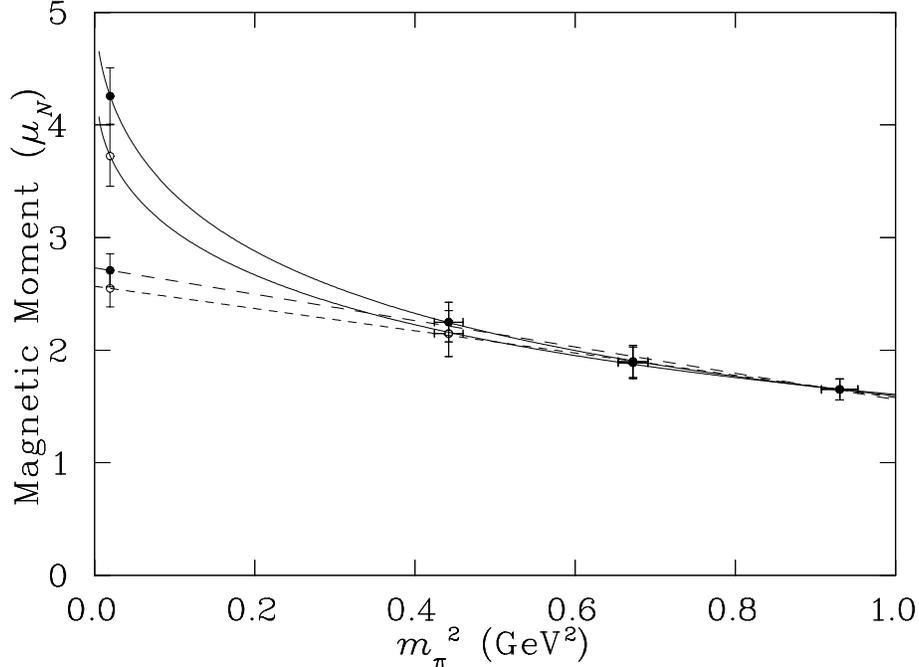,height=12cm}} }
\caption{Extrapolation of the valence $u$-quark magnetic moment
contributions to the proton (solid symbols) and $\Sigma^+$ (open
symbols).  Naive linear extrapolations (dashed lines) are contrasted
with the chiral extrapolations based on (\protect\ref{eq:chiral}).
The lattice data is taken from 
Ref.\protect\cite{dblOctet}.
}
\label{PSigExtrap}
\end{figure}

Figures \ref{NXiExtrap} and \ref{PSigExtrap} illustrate the
extrapolations of the lattice data.  The results of these fits shown
at $m_\pi = 140$ MeV are the results of the extrapolation procedure,
including fitting errors.  Naive linear extrapolations are also shown
to help emphasize the dominant role of chiral symmetry in the
extrapolation process.

Clearly the effect of the chiral corrections on the extrapolated values
of the key magnetic moment ratio
\begin{equation}
%\frac{S_N}{S_\Xi} \equiv 
\frac{\mu_u^n(m_\pi = 140 MeV)}{\mu_u^{\Xi^0}(m_\pi=140MeV)}
\label{nXiratio}
\end{equation}
is dramatic. Instead of the value $0.72 \pm 0.46$ obtained by linear
extrapolation one now finds $1.51 \pm 0.37 $.  By comparison, the
ratio
\begin{equation}
\frac{\mu_u^p(m_\pi = 140 MeV)}{\mu_u^{\Sigma^+}(m_\pi = 140 MeV)}
\end{equation}
is quite stable, changing from $1.14 \pm 0.08\ $ in the case of linear
extrapolation to $1.14 \pm 0.11 $ when the correct chiral
extrapolation is used.  In lattice terminology the ${u_n}/{u_{\Xi^0}}$
case involves a ``singly represented'' quark, whereas
${u_p}/{u_{\Sigma^+}}$ involves the ``doubly represented'' quark. It
is perhaps not too surprising that the breakdown of the universality
of the effective quark magnetic moments should be bigger for the
quarks which are in the minority.  It is vital that the ratio obtained
in Eq.(\ref{nXiratio}) is much more consistent with the range of
values found necessary (in Sect. 2) to yield a positive value of
$G_M^s$, as illustrated in Fig.\ \ref{SelfCons} -- values which seemed
quite unreasonable in the constituent quark model.

\begin{figure}[tbp]
\centering{\
\rotate{\epsfig{file=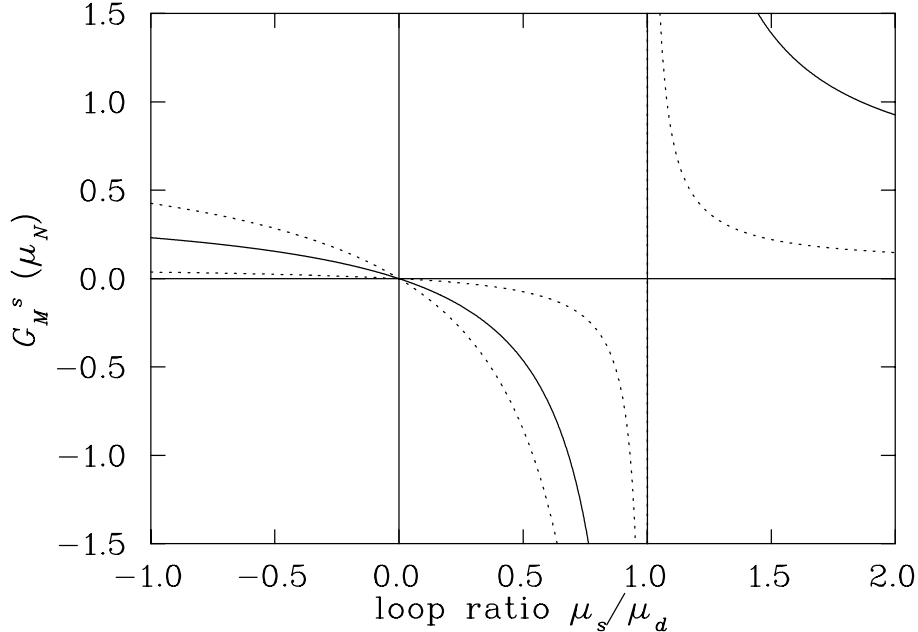,height=12cm}} }
\caption{The strange quark contribution to the nucleon magnetic moment
$G_M^s(0)$ obtained from (\protect\ref{ok}) plotted as a function of
$^lR^s_d$ reflecting the relative contribution of strange to light
sea-quark loop contributions.  The solid curve illustrates the central
values and the dashed curves denote the standard error.  }
\label{GMsRsdSigmaP}
\end{figure}

\begin{figure}[tbp]
\centering{\
\rotate{\epsfig{file=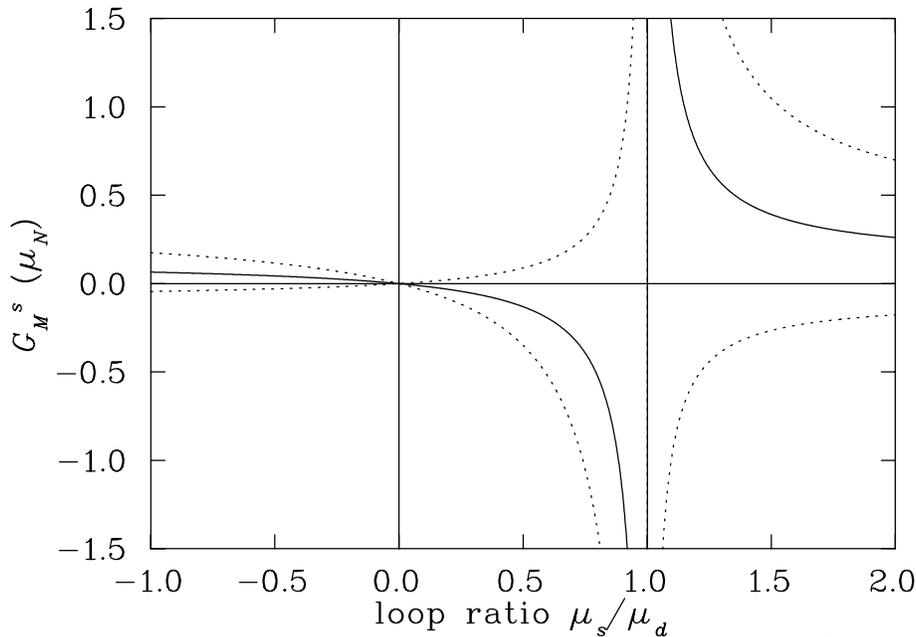,height=12cm}} }
\caption{The strange quark contribution to the nucleon magnetic moment
$G_M^s(0)$ obtained from (\protect\ref{great}) plotted as a function of
$^lR^s_d$ reflecting the relative contribution of strange to light
sea-quark loop contributions.  The solid curve illustrates the central
values and the dashed curves denote the standard error.  }
\label{GMsRsdXiN}
\end{figure}

Figures \ref{GMsRsdSigmaP} and \ref{GMsRsdXiN} illustrate the allowed
range of $G_M^s$, using (\ref{ok}) and (\ref{great}) respectively,
with these new valence ratios (as a function of the loop ratio
$^lR^s_d$).  At the preferred value, $^lR^s_d = 0.65$, (\ref{ok}) and
(\ref{great}) yield $G^s_M = -0.57 \pm 0.42 \mu_N$ and $-0.16 \pm 0.18
\mu_N$, respectively.

As we have already remarked, the effect of the chiral extrapolation on
the ratio ${u_n}/{u_{\Xi^0}}$ is the most dramatic, changing it from
$0.72 \pm 0.46$ to $1.51 \pm 0.37$. Thus Eq. (\ref{great}), which
seemed certain to guarantee $G_M^s < 0$, no longer does. Indeed, for
the preferred value $^lR^s_d$ = 0.65, we obtain $G^s_M = -0.16 \pm
0.18 \mu_N$. While this is still negative it does permit a small
positive value within the error.

\section{Summary Remarks and Outlook}

We have presented two equations, (\ref{ok}) and (\ref{great}),
describing the strange quark contribution to the nucleon magnetic
moment, $G_M^s(0)$, in terms of the ratio of strange to light
sea-quark-loop contributions and valence-quark ratios which probe the
effects of environment sensitivity.  These equations are {\em exact}
consequences of QCD under the assumption of charge symmetry (i.e.,
$m_u = m_d$ and negligible electromagnetic corrections).

The sea-quark-loop ratio probes the quark-mass suppression of the
strange-quark loop relative to the light-quark loop such that the
ratio is expected to lie between 0 and 1.  Traditionally this ratio is
given by a ratio of constituent quark masses yielding 0.65.  

The valence-quark ratios probe the effects of environment sensitivity.
In this discussion we have seen how differences in the pion cloud
contributions to various baryons give rise to very significant
environment sensitivity in the quark sector contributions to baryon
magnetic moments.  Indeed the environment sensitivity in the $u$-quark
contributions to $n$ versus $\Xi^0$ are so large,
(${u_n}/{u_{\Xi^0}} = 1.51 \pm 0.37$), that one should seriously
reconsider the validity of a constituent quark picture for the
minority, or ``singly represented'' valence quarks.  Using the new
value for ${u_n}/{u_{\Xi^0}}$ leads to a strangeness magnetic moment
for the proton, $G_M^s(0)$, equal to $-0.16 \pm 0.18 \mu_N$.

The lattice QCD simulation results themselves display an environment
sensitivity.  For example, the strange quarks in the $\Xi^0$ act to
enhance the magnitude of the $u$-quark contribution to the $\Xi^0$
moment relative to that in the neutron in the simulation results.
However, this environment sensitivity may be an overestimate, as
the mass of the strange quark used in the lattice QCD simulations is
somewhat heavy, approaching 300 MeV.  

As an estimate of the effects of this systematic error, we may average
the neutron and $\Xi^0$ simulation results for the $u$-quark to
linearly interpolate to a strange quark mass the order of 150 MeV.  In
this case the magnetic moment ratio, $u_n/u_{\Xi^0}$, increases to
1.9(6) providing a positive value for $G_M^s(0)$ at $0.09(22)\ \mu_N$,
a shift of 0.25 $\mu_N$ in the central value.  Similarly the
$u_p/u_{\Sigma^+}$ ratio becomes 1.09(10) and provides $G_M^s(0) =
-0.35(39)\ \mu_N$, also consistent with small positive values for
$G_M^s(0)$, within one standard deviation.  In the absence of new
lattice calculations with a realistic strange quark mass, it is not
possible to make a better estimate of the systematic error on
$G_M^s(0)$.  We find it remarkable that the approach described here
can produce a reasonably precise estimate of $G_M^s(0)$ from the
$\Xi^0$ and $n$ systems even with lattice data that is quite old.
Advances in high performance computing and lattice QCD methodology in
the last seven to eight years mean that it should be possible to
dramatically reduce both the statistical and systematic errors.  In
view of the present results, and the importance of understanding the
role of strangeness in the nucleon, this is clearly an urgent priority
-- as is the need to obtain unquenched magnetic moment results at
masses as low as possible.

\section*{Acknowledgments}
We would like to acknowledge very helpful discussions concerning the
work described here and particularly this manuscript with Pierre Guichon
and Tony Williams.
This work was supported by the Australian Research Council.

%\section*{References}

%\bibliography{../../../biblio/biblio} %,GaussSR}
% below {nh} for North-Holland, {aip} for AIP journals 
%\bibliographystyle{nh}

\end{document}